\documentstyle[12pt]{article}
\begin{document}
\def\om{\omega}
\def\omt{\tilde{\omega}}
\def\ti{\tilde}
\def\o{\Omega}
\def\t{T^*M}
\def\vt{\tilde{v}}
\def\ot{\tilde{\Omega}}
\def\otwo{\omt \wedge \om}
\def\owot{\om \wedge \omt}
\def\w{\wedge}
\def\mt{\tilde{M}}

\def\om{\omega}
\def\omt{\tilde{\omega}}
\def\ss{\subset}

\def\om{\omega}
\def\omt{\tilde{\omega}}
\def\ti{\tilde}
\def\o{\Omega}
\def\t{T^*M}
\def\vt{\tilde{v}}
\def\ot{\tilde{\Omega}}
\def\otwo{\omt \wedge \om}
\def\owot{\om \wedge \omt}
\def\w{\wedge}
\def\mt{\tilde{M}}

\def\om{\omega}
\def\omt{\tilde{\omega}}
\def\ss{\subset}
\def\tpm{T_{P} ^* M}
\def\al{\alpha}
\def\alt{\tilde{\alpha}}
\def\la{\langle}
\def\ra{\rangle}
\def\inop{{\int}^{P}_{P_{0}}{\om}}
\def\th{\theta}
\def\tht{\tilde{\theta}}
\def\inox{{\int}^{X}{\om}}
\def\inotx{{\int}^{X}{\omt}}
\def\st{\tilde{S}}
\def\ls{\lambda_{\sigma}}
\def\p{{\bf{p}}}
\def\pb{{\p}_{b}(t,u)}
\def\pbm{{\p}_{b}}
\def\d{\partial}
\def\d+{\partial_+}
\def\d-{\partial_-}
\def\pat{\partial_{\tau}}
\def\pas{\partial_{\sigma}}
\def\dpm{\partial_{\pm}}
\def\l2{\Lambda^2}
\def\be{\begin{equation}}
\def\ee{\end{equation}}
\def\bea{\begin{eqnarray}}
\def\eea{\end{eqnarray}}
\def\ej{{\bf E}}
\def\ed{{\bf E}^\perp}
\def\si{\sigma}
\def\cg{{\cal G}}
\def\cgt{\ti{\cal G}}
\def\cd{{\cal D}}
\def\ce{{\cal E}}
\def\cep{\ce^{\perp}}
\def\cf{{\cal F}}
\def\cfp{\cf^{\perp}}
\def\bz{\bar{z}}
\def\e{\varepsilon}
\def\b{\beta}
\begin{titlepage}
\begin{flushright}
{}~
CERN-TH/96-43\\
hep-th/9602162
\end{flushright}

\vspace{1cm}
\begin{center}
{\Large \bf  Dressing Cosets}\\
[50pt]
{\small
{\bf C. Klim\v{c}\'{\i}k}\\
Theory Division CERN \\
CH-1211 Geneva 23, Switzerland \\[10pt] and \\[10pt] 
{\bf P. \v Severa }\\
Department of Theoretical Physics, Charles University, \\
V Hole\v sovi\v ck\'ach 2, CZ-18000 Praha,
Czech Republic\\[40pt] }

\begin{abstract}
 The account of the Poisson-Lie T-duality is presented for the case 
when the
action of the duality group on a target is not free. 
At the same time 
a generalization of the picture is given when the duality group
does not even act on $\si$-model targets but only on their phase spaces.
The outcome is
a huge class of dualizable targets generically having  no local isometries
or Poisson-Lie symmetries whatsoever.

\end{abstract}
\end{center}
\vskip 2.5cm
\noindent CERN-TH/96-43\\
February 1996

\end{titlepage}

\noindent 1. The Poisson-Lie (PL) $T$-duality \cite{KS2} is the 
generalization of the 
traditional non-Abelian $T$-duality \cite{OQ}--\cite{GRV}. It has been
demonstrated in \cite{KS2} and in the  series of subsequent papers
\cite{T} -- \cite{KS4} that the PL $T$-duality enjoys most of the
structural features of the traditional Abelian $T$-duality 
\cite{SS} -- \cite{GPR}.

 The underlying structure of the PL $T$-duality is the Drinfeld double
\cite{D}. The latter is the Lie group which is a sort of twisted product 
of two
its equally dimensional subgroups. These subgroups play the role of the
duality and coduality groups in the following sense: 
The duality group acts on the target of a PL dualizable $\sigma$-model
and this action is Poisson-Lie symmetric with respect to the 
coduality group (see \cite{KS2} for the definition
of the PL symmetry). In the dual $\si$-model
the roles of the duality and coduality groups are interchanged.

 In the
traditional non-Abelian duality the duality group is some Lie group $G$ 
and coduality group is its Lie coalgebra viewed as the commutative additive 
group.
The Drinfeld double is the cotangent bundle of the group manifold $G$ in this
case.

It has been remarked already many times \cite{OQ,GR,GRV} that even in the
framework of the traditional non-Abelian duality there is the  
 possibility of a qualitatively new structure which is absent in the Abelian
case. It is connected with the fact that a non-Abelian duality group may
act with isotropy which means, in other words, that the action is not free.
A concrete example of the non-Abelian dual model in the case
of the non-free action of the duality group was worked out e.g. in 
\cite{OQ,GR,Alv1,Hew}
by the standard method of gauging of isometry. 

The purpose of this note is to generalize the results of the traditional
non-Abelian duality for the not freely acting groups to the general
Poisson-Lie case and to find the relevant algebraic structure in terms
of the corresponding Drinfeld double. We  find that in this case 
the PL duality relates $\si$-models on the targets which are respectively
cosets of an appropriate (dressing) action of certain residual group on the
duality and coduality groups. In general, there is no action of
the duality or the coduality group on these cosets,
and, consequently, no trace of isometry or Poisson-Lie symmetry
of the targets\footnote{These `dressing' cosets $\si$-models 
should presumably fit well into the schemes of \cite{A,EG}.}. Still the 
duality and the coduality group underlie
the dynamics of the $\si$-models in a non-local way.

In what follows, we  give the duality invariant
description of a Hamiltonian dynamical system on the subspace
of the loop group of the Drinfeld double and show that this system
 simultaneously
describes the dynamics of the both coset $\si$-models related by 
the PL duality. Then we discuss  concrete examples of the traditional 
non-Abelian $T$-duality and of the `true' PL duality with both the duality 
and the coduality groups
being non-Abelian.
\vskip1pc
\noindent 2. 
For the description of the Poisson-Lie duality we need the
crucial concept
of the Drinfeld double  which is simply  a  Lie group $D$ such that
its Lie algebra
$\cd$ can be decomposed into a pair of maximally isotropic subalgebras with
respect to a non-degenerate invariant bilinear form on $\cd$ \cite{D}.

Consider now an $n$-dimensional linear subspace $\ce$ of the $2n$-dimensional
Lie algebra
$\cd$ and its orthogonal complement $\ce^{\perp}$ such that the intersection
$\ce \cap \ce^{\perp}\equiv\cf$ is an isotropic Lie subalgebra of $\cd$. 
Moreover
we require that the both subspaces $\ce$ and $\cep$ 
are invariant with respect to the adjoint
action of $\cf$. 
We shall show that these
data determine a dual pair of  $\si$-models with the targets being
the dressing cosets of the groups  $G$ and $\ti G$ respectively. These cosets 
are defined with respect to the dressing action of the group $F$ whose Lie 
algebra is $\cf$. The dressing action of an element $f\in F$ on an element
$g\in G$ gives an element $g_1\in G$ defined as follows
\be fg=g_1\ti h, \quad \ti h\in \ti G.\ee
The multiplication in (1) is understood in the sense of the Drinfeld 
double\footnote{The element $g_1$ is well defined if $f$ and $g$ are close 
to unit
and for some Drinfeld doubles the definition is entirely correct even globally.
For a generic double a special global analysis is required which, however,
 does not
elucidate the main idea of the note and is in fact beyond the scope of it.}.
By the dressing coset we mean the set of orbits of the dressing action
of $F$ on $G$ or $\ti G$.

The most economic description of the common dynamics of the models from
the dual pairs is given in terms of the loop group $LD$ of the Drinfeld double.
The phase space $P$ is formed by the loops $l(\si)$ with the property
\be \partial_{\si} l l^{-1} \in \cfp,\ee
where $\cfp$ denoted the orthogonal complement of $\cf$ with respect to the
invariant inner product on the double. Note that $\cf$ is isotropic, hence
$\cf\ss\cfp$.
 We also postulate that the loops
$l_1(\si)$ and $l_2(\si)$ such that
\be l_1(\si)=l_2(\si) l_c , \quad l_c\in D \ee
 by definition describe the
same element of the phase space. Note that the right action of $D$
leaves invariant the current component $\partial_{\si} l l^{-1}$.

We define a symplectic two-form $\Omega$ on this phase space
as the exterior derivative of a  polarization one-form $\alpha$. The latter
is most naturally defined in terms of its integral along an arbitrary
curve $\gamma$ in the phase space, parametrized by a parameter $\tau$.
From the point of view of the Drinfeld double, this curve is a surface with
the topology of cylinder embedded in the double in such a way that 
constraint (2) is fulfilled.  We define
\be \int_{\gamma} \alpha ={1\over 2}\int \la \pas l~l^{-1},
\pat l~l^{-1}\ra+{1\over 12} \int d^{-1}\la dl~l^{-1},[dl~l^{-1},
dl~l^{-1}]\ra .\ee
Here $\la .,.\ra $ denotes the non-degenerate invariant bilinear form
on the Lie algebra ${\cal D}$  of the double and in  the second term
 on the r.h.s. we
recognize
the two-form
potential
of the WZW three-form on the double. Note that this definition of $\alpha$ is
ambiguous due to the ambiguity in the choice of 
the inverse exterior derivative  $d^{-1}$. However, this ambiguity disappears
when the exterior derivative of the one-form $\alpha$ is taken. In other words,
the symplectic form $\Omega$ is well defined.

We should note that we use the notion of symplectic
form somewhat loosely. By this we mean that the symplectic form
$\Omega$ is closed but it is {\it not } non-degenerate.
From the point of view of the Hamiltonian mechanics this corresponds to 
the situation occuring in the description of systems with gauge symmetry.
The vector fields annihilating $\Omega $ (i.e. $\Omega(.,v)\equiv 0$) form
an algebra under the standard Lie bracket hence they give rise to integrable
surfaces (=orbits of the gauge group) in $P$ on which $\Omega$ identically 
vanishes. By factoring the original phase space by these gauge
group orbits, we obtain a reduced phase space on which $\Omega$ is not
only closed but is also non-degenerate. If we define
a Hamiltonian on the original phase space which is (gauge) invariant with 
respect to the action of those vector fields we have a well defined 
Hamiltonian
system on the reduced phase space.

In our concrete situation, we define a Hamiltonian in terms of a certain
quadratic form $K$ on $\cfp$ such that
\be K(x+x_0)=K(x) , \quad x\in \cfp , x_0 \in \cf.\ee
The value of K on some vector $x\in \cfp$ is computed as follows: $x$
can be (not uniquely) decomposed as 
\be x=x_1+x_2,  \quad x_1\in \ce, \quad  x_2\in \cep\ee
and 
\be K(x)\equiv \la x_1,x_1\ra-\la x_2,x_2\ra.\ee
Note that  the value $K(x)$ does not depend on the
decomposition (6).

In this note, we shall study a dynamical system on the phase space $P$
given by the action
$$ S[l(\tau,\si)]=\int \alpha - \int H~d\tau $$
\be =\int d\si d\tau \big\{{1\over 2}\la \pas l~l^{-1},\pat l~l^{-1}\ra+
{1\over 12}
d^{-1}\la dl~l^{-1},[dl~l^{-1},
dl~l^{-1}]\ra -{1\over 2}K(\pas l l^{-1})\big\}.\ee
It is easy to check that the  group action corresponding to the vector
fields annihilating
the symplectic form $\Omega$ is given by the left multiplication
of a loop $l(\si)$ by an element $f(\si)$ from the loop group $LF$.
The Hamiltonian $H$
is invariant with respect to this action. 

We conclude that  the data $P$, $\Omega$
and $H$ yield a well-defined Hamiltonian system on the reduced phase space
$LF\backslash P$. The description of this system in terms of the original
 phase space
$P$ is given by the first order action $S$ which, as it should,
 indeed possesses
the gauge symmetry with respect to the left multiplication of $l(\si,\tau)$
by arbitrary $f(\si,\tau)\in F$:
\be l(\si,\tau)\to f(\si,\tau) l(\si,\tau). \ee
 Note that the action $S$ has also 
a little gauge invariance corresponding to the write multiplication
of $l(\si,\tau)$ by arbitrary function $l(\tau)\in D$. This small gauge
symmetry corresponds to the factorization (3).

\vskip1pc
3. We show that the Hamiltonian system, defined by $P$, $\Omega$ and $H$,
simultaneously describes dynamics of two $\si$-models. Their Lagrangians 
may be obtained directly from the action (8) as follows:
Write the field $l(\si,\tau)$ in the form
\be l(\si,\tau)=g(\si,\tau) \ti h(\si,\tau).\ee
Here $g(\si,\tau)$ is an unconstrained element of $G$ and $\ti h(\si, \tau)$
is from $\ti G$ in such a way that $l(\si,\tau)$ fulfils the constraint (2).
Now $\ti h$ can be eliminated from the action (8), yielding the $\si$-model
on the group manifold $G$ with the Lagrangian

\be L=(E+\Pi(g))^{-1}(\partial_+ g g^{-1}, \partial_- g g^{-1}).\ee
Here the indices $\pm$ mean the light cone variables on the world
sheet and  $E$ is a  bilinear form
  on the dual space $\cg^*$ of
the Lie algebra $\cg$ of the group $G$. The graph of $E$ in $\cd$ is 
precisely the subspace $\ce$, i.e.
\be \ce=Span\{t+E(t,.),t\in\cgt\},\quad  \cep=
Span\{t-E(.,t),t\in\cgt\}.\ee
$\Pi(g)$ is the bivector field
on the group manifold  which  gives the  Poisson-Lie bracket on $G$ 
(i.e. the multiplication $G\times G\to G$ is the Poisson map) 
\cite{D, KS2,KS3}.

By choosing the dual ansatz
\be  l(\si,\tau)=\ti g(\si,\tau)  h(\si,\tau)\ee
we arrive at the dual $\si$-model on the dual group $\ti G$ manifold:
\be \ti L=(E^{-1}+\ti \Pi(\ti g))^{-1}(\partial_+ \ti g \ti g^{-1},
\partial_- \ti g \ti g^{-1}).\ee
The mutually dual $\si$-models (11) and (14) appear to live on the targets
$G$ and $\ti G$ respectively but, in fact, they do not.
The standardly computed symplectic forms on their phase spaces are
degenerate. The reason
is the gauge symmetry (9) of the original model (8). Therefore
the  resulting $\si$-models (11) and (14) possess the same gauge 
symmetry but now the group $F$ acts from the left not by the 
standard multiplication as in (9) but by the dressing action (1).
Thus the $\si$-models (11) and (14) live on the targets which are respectively
cosets of the dressing action of the group $F$ on $G$  and on $\ti G$.

In every concrete example we may choose  convenient gauge slices cutting
the  orbits  of the dressing action and
to work out the  targets of the $\si$-models (11) and (14) in terms 
of some coordinates on the slices. We shall do it explicitly in some
cases in order to illustrate the method. 

It is interesting to note
that there is no natural action of the duality group $G$ on the gauge
fixed model (11) . The only exception occurs when
  $F$ is a  subgroup of 
the  group $G$ . In this case the target $F\backslash G$ is the
standard coset on which $G$ naturally acts. The isotropy subgroup
of this action is precisely $F$ and we recover the 
standard picture of the traditional non-Abelian duality.
 But also in this special case, $F$ is
not subgroup of $\ti G$ and therefore there is no natural action of $\ti G$
on the dual $\si$-model target.

The suggested derivation of the gauge-invariant $\si$-models (11) and (14)
from the duality invariant action (8) is technically quite lengthy. 
It is easier to demonstrate the equivalence of (8), (11) and (14)
by a short-cut argument. For concreteness, let us consider the equivalence
of the models
(8) and (11). The density of the canonical momentum of (11)
is a one-form with
values  in the coalgebra
$\cg^*$. On the other hand, 
$\cg^*$  is canonically identified with $\cgt$ by means
of the invariant inner product in $\cd$.
It turns out (see \cite{KS2,KS3} for a detailed argument) that 
the density of the canonical momentum on an extremal configuration
$g(\si,\tau)$ can be written as  the zero-curvature form 
$d\ti h \ti h^{-1}$ for some $\ti h(\si,\tau)\in \ti G$.  Hence for every
extremal configuration of the model (11) or, in other words, at  every
point of the phase space of (11), we may find a configuration $l(\si,\tau)$
in the double given by 
\be l=g\ti h.\ee
This configuration is determined up to the right
 multiplication by a constant element from $\ti G$. 

It is now very easy to check that under the mapping
(15) the standard
polarization form ($pdq$) for the $\si$-model (11) coincides with
the polarization form $\alpha$ given in (4) (for a specific choice of
$d^{-1}$). Moreover, the Hamiltonian of (11) also coincides 
with the Hamiltonian of (8). Thus we conclude that the models (8) and (11)
(and in the same way (8) and (14)) are dynamically equivalent. In fact,
it is much easier to study the (dressing)
gauge invariance of the models (11) and (14)
in terms of the standard gauge invariance (9) of the
duality invariant action (8), where the  
invariance of the symplectic form and of the Hamiltonian is manifest.

\vskip1pc
4. Now there is time for some examples. The simplest one is the  sphere
with the round metric and invariant 2-form, dualized with respect to $SU(2)$.
The double is the cotangent bundle of $SU(2)$, algebra $\cf$ is generated by
the Pauli matrix ${\si}_3$, $\ce =Span(\si _3,\si _+ +ia t_+,\si _- -ib t_-)$,
where $t_i$ is the basis of the coalgebra of $SU(2)$ dual to $\si _i$  and 
$a,b$ are arbitrary real parameters. The 
result is the standard round metric on  $F\backslash SU(2)$ (=2-sphere)
 and the standard monopole 2-form as the torsion: 
\be{{a-b}\over 2}(d\theta ^2+\sin ^2 \theta d\phi ^2), \quad
 {{a+b}\over 2}\sin \theta
d\theta \w d\phi . \ee
Now the dressing action of the group $F$ on the coalgebra is simply rotation
with respect to the $z$-axis,  the torsion 2-form vanishes and the dual metric
on the dressing coset (having the topology of the half-plane) reads
\be{1 \over {(b-a)\rho}}(d\rho+(z-a)dz)(d\rho+(z-b)dz).\ee
Here $\rho$ is one half of the squared distance from the $z$-axis.

So far we have rederived the  result of the traditional non-Abelian T-duality
\cite{OQ,GR,Alv1,Hew}.
Now we present its generalization, when the cotangent bundle is replaced by 
$SL(2,C)$ and the coalgebra is replaced by the Borel group $B_2$ of 
upper-triangular matrices in
$SL(2,C)$ with real entries on the diagonal. The invariant bilinear form
on the double is $<a,b>={1\over \epsilon} Im(tr(ab))$ with an arbitrary
real $\epsilon$. $\cf$ and $\ce$ remain the same, written in terms of the
elements of the original basis $\si_i$ and its dual basis
$t_i$ (now defined with respect to the inner product on $sl(2,C)$).
 The metric and the torsion 2-form
on the sphere are
\be {1 \over \Delta}{{a-b}\over 2}(d\theta ^2+\sin ^2 \theta d\phi ^2), \quad
{1 \over \Delta}({{a+b}\over 2}+2\epsilon a b \sin ^2 {\theta \over 2}) 
\sin \theta d\theta \w d\phi,\ee
$$\Delta=(1+2\epsilon a \sin ^2 {\theta \over 2})(1+2\epsilon b 
\sin ^2 {\theta \over 2}).$$
 The dual torsion 2-form vanishes and the dual metric in appropriate 
coordinates reads
\be {1\over {1+\epsilon z}}{1\over {(b-a)\rho}}
(d\rho+({{z+\epsilon z^2 /2}\over {(1+\epsilon z)^2}}-a)dz)
(d\rho+({{z+\epsilon z^2 /2}\over {(1+\epsilon z)^2}}-b)dz).\ee

Note that in the limit $\epsilon\to 0$ our $SL(2,C)$ results
(18) and (19) reproduce the traditional non-Abelian duality results
(16) and (17). Thus we have obtained a one-parametric deformation
of the dual pair of \cite{OQ,GR,Alv1,Hew}. 

 The data (16) and (18) are defined on the standard
coset $F\backslash SU(2)$ where the duality group $SU(2)$ naturally acts.
Only the data (17) and (19) are defined on the truly dressing
cosets where there is no natural action of the $SU(2)$ coalgebra 
and the Borel group $B_2$            on the coset targets (17) and (19)
respectively. It is not too difficult to find examples, where   the both 
targets
from the dual pair are truly dressing cosets. The corresponding formulas
in explicit
coordinates 
are not very illuminating, however.

  \vskip1pc
5. We conclude that there is the natural generalization of the
traditional non-Abelian $T$-duality with a non-freely acting duality group.
In the most general case the duality group does not even act on the
$\si$-model target but in the non-local way on its phase space.
Only in the special case when
 the residual group $F$ is a subgroup of the duality group $G$  
the action of $G$ on the $\si$-model target is local and $F$ is the isotropy
group of this action. We should mention that the global aspects of the PL 
$T$-duality \cite{KS4} can be also settled in the case of the dressing cosets.
We have shown in \cite{KS4} that the basic
data defining the PL $T$-duality between D-branes are the $2n$-dimensional
Drinfeld double
and $n$-dimensional isotropic subalgebra ${\cal A}$ of the Lie algebra of 
the double. 
If our algebra $\cf$  is also the subalgebra of ${\cal A}$ then all results
of \cite{KS4} directly generalize to the dressing cosets.

There remains an  important nontrivial open problem: Is a given $\si$-model a 
dualizable dressing coset? The nontriviality stems from the fact
that even if the answer is in some cases
affirmative the duality group does not act on the target and the dressing
orbits are in general too `wild' to be easily recognizable. On the other
hand, we find particularly this aspect of our construction promising.
The simple data on the double give rise to very non-symmetrically looking
$\si$-models whose targets, metrics and torsions are straightforwardly
 defined but not easily evaluated explicitly. Needless to say, 
eventually   we hope to establish a connection of the PL $T$-duality with
the mirror symmetry.

Another interesting project consists in considering
the subspace $\ce$ to be an isotropic subalgebra. The resulting cosets
should be  topological theories and the PL duality would rotate just
the zero modes in a nontrivial way \cite{AKS}. At the  quantum level,
a path integral derivation of the dressing cosets should be obtained perhaps
by a modification of the derivation due to Tyurin and von Unge  \cite{TU} 
or in the way suggested in \cite{KS3}.

\vskip 2pc
We thank A. Alekseev and E. Kiritsis for discussions.

\end{document}